\documentclass[12pt]{article}
\usepackage{color}
\definecolor{myblue}{rgb}{.8, .8, 1}

\usepackage{amsmath}   
\usepackage{empheq}
\usepackage{amsfonts}
\usepackage[bbgreekl]{mathbbol}
\makeatletter
\def\amsbb{\use@mathgroup \M@U \symAMSb}
\makeatother
\usepackage{amssymb,amsfonts}

\usepackage{graphicx}

\newlength\mytemplen
\newsavebox\mytempbox

\makeatletter
\newcommand\mybluebox{%
    \@ifnextchar[%]
       {\@mybluebox}%
       {\@mybluebox[0pt]}}

\def\@mybluebox[#1]{%
    \@ifnextchar[%]
       {\@@mybluebox[#1]}%
       {\@@mybluebox[#1][0pt]}}

\def\@@mybluebox[#1][#2]#3{
    \sbox\mytempbox{#3}%
    \mytemplen\ht\mytempbox
    \advance\mytemplen #1\relax
    \ht\mytempbox\mytemplen
    \mytemplen\dp\mytempbox
    \advance\mytemplen #2\relax
    \dp\mytempbox\mytemplen
    \colorbox{myblue}{\hspace{1em}\usebox{\mytempbox}\hspace{1em}}}

\makeatother

\usepackage{bbm}
\usepackage{yfonts}
\pdfoutput=1
\usepackage{amsfonts}
\usepackage[bbgreekl]{mathbbol}
\makeatletter
\def\amsbb{\use@mathgroup \M@U \symAMSb}
\makeatother
\usepackage[bbgreekl]{mathbbol}
\usepackage{amssymb,amsfonts}
\usepackage{graphicx}
\usepackage[top=2.75cm, bottom=2.75cm, left=2.75cm, right=2.75cm]{geometry} 
\usepackage{appendix}
\usepackage{color}
\usepackage{caption}
\usepackage{lipsum}

\newcommand{\be}{\begin{equation}}
\newcommand{\ee}{\end{equation}}
\newcommand{\ber}{\begin{eqnarray}}
\newcommand{\eer}{\end{eqnarray}}

\def\+{{+\!\!\!+}}

\newcommand{\nn}{\nonumber}
\newcommand{\pa}{\partial}
\newcommand{\na}{\nabla}
\newcommand{\half}{{\textstyle{\frac12}}}

\newcommand{\bbD}[1]{\mathbb{D}_{#1}}
\newcommand{\bbDB}[1]{\bar{\mathbb{D}}_{#1}}

\newcommand{\re}[1] {(\ref{#1})}

\numberwithin{equation}{section}

\begin{document}
\begin{titlepage}
\begin{flushright} \small
Imperial-TP-2017-CH-02\\
UUITP-20/17\\
%Imperial-TP-2017-UL-01\\
\
\end{flushright}
\smallskip
\begin{center} 
\LARGE
{\bf  All $(4,0)$:   
 Sigma Models with $(4,0)$ Off-Shell Supersymmetry}
%All Four Nothing}
 \\[30mm] 
\large
{\bf Chris Hull$^a$}~and~{\bf Ulf~Lindstr\"om$^{ab}$} \\[20mm]
{ \small\it
$^a$The Blackett Laboratory, Imperial College London\\
Prince Consort Road London SW7 @AZ, U.K.\\
\vspace{2mm}
$^b$ Department of  Physics and Astronomy,
\\ Division of Theoretical Physics,
Uppsala University,\\ Box 516, SE-751 20 Uppsala, Sweden }
\end{center}

\vspace{10mm}
\centerline{\bfseries Abstract} 
\noindent
%Off-shell $(4,0)$ supermultiplets in  2-dimensions are constructed 
Off-shell $(4,0)$ supermultiplets in  2-dimensions are formulated. These are used to construct sigma models whose target spaces are vector bundles over manifolds that are  hyperk\"ahler with torsion. The off-shell supersymmetry implies that the complex structures are simultaneously integrable and allows us to  write  actions using extended   superspace and projective superspace, giving an explicit construction of the target space geometries.
\bigskip

\end{titlepage}

\tableofcontents
\section{Introduction}

The     $(1,0)$   supersymmetric sigma model  \cite{Hull:1985jv}  
 has a target space ${\cal M}$ with  a metric $g$ and closed 3-form $H=dB$, together with a vector bundle over ${\cal M}$
with connection $A_i {}^B{}_C(X)$ and fibre metric $G_{AB}(X)$, where $A,B=1,\dots ,m$ 
are indices for the $m$-dimensional fibres. 
The action can be written in $(1,0)$   superspace with coordinates 
$x^\mu = (x^\+,x^=)$ and $ \theta^-$.
The action is given in terms of scalar superfields $X^i(x,\theta)$ with  $i=1,...,D$ (where $D$ is the real dimension of the target space)
and $m$ real fermionic spinor superfields $\Psi^A(x,\theta)$ with $A=1,\dots , m$. 
%If the target space coordinates are $X^i$, $i=1,...,n$, the map from the worldsheet superspace to the target %space is given locally by scalar superfields $X^i(x,\theta)$ and t
The action is %
\ber \label{heter}
S=\frac 1 2 \int d^2x 
d \theta \,\left(
\partial _= X^i (g+B)_{ij}D_+X^j +G_{AB}\Psi^A _-\nabla_+\Psi ^B   _-
\right)
\eer
where
\ber
\nabla_+\Psi ^B   =D_+\Psi ^B +D_+ X^i A_i {}^B{}_C \Psi ^C
\eer
Only the antisymmetric part of the connection appears in the action so we can take
$A_{iAB}  \equiv G_{AB}A_i {}^B{}_C$ to be antisymmetric, $A_{iAB}=-A_{iBA}$, and the structure group $G$ of the bundle  to be in $O(m)$.
The modified connection
\ber
\hat A_i {}^B{}_C = A_i {}^B{}_C + \frac 1 2 G^{BA}G_{AC,i}
\eer
is a metric connection, $\hat \nabla _i G_{AB}=0$ \cite{Hull:1985jv}. As replacing $A$ with $\hat A$ in the action {\re{heter}} doesn't change the action (the difference between the two actions vanishes),  we can 
without loss of generality take  $A$ to be the metric    connection  \cite{Howe:1988cj}, and we drop the hat from $A$ in what follows.

The classical $(1,0)$ sigma model will in fact be $(p,0)$ supersymmetric provided that ${\cal M}$ admits $p-1$ complex structures $J^{(A)}$  such that the geometry satisfies the conditions~\cite{Hull:1986hn}
 
% for {$(p,1)$} supersymmetry,  {i.e., 
\ber
J^{(A)t}gJ^{(A)}=g~,~~~(J^{(A)})^2=-\mathbb{1}~,~~~\na^{(+)}J^{(A)}=0~,
\eer
where 
\ber
\na^{(+)}:=(\na^{(0)}+\half g^{-1}H)
\eer
is the connection with torsion $g^{il}H_{ljk}$ added to the  Levi-Civita connection
$\na^{(0)}$.
%discussed above, 
 {In addition it is required }%that together with the condition 
 that the vector bundle is holomorphic with respect to each of the complex structures, i.e the field strength
$F=dA +\frac 1 2   [A,A]$ is a (1,1) form with respect to each of the complex structures \cite{Hull:1985jv}, \cite{Howe:1988cj},\cite{Hull:1986hn}. 
The supersymmetry transformation  { of the bosonic superfields is 
\ber\label{nis1tfs}
\delta X^i=\epsilon_A^+\left(J_{(A)}\right)^i_jD_+X^j~,
\eer
and that of} the fermi superfields is
\be \label{psitrans}
\delta \Psi ^B  =-  (\delta X^i ) A_i {}^B{}_C \Psi ^C~.
\ee
%where $(\delta X^i )$ is given by
%\re{nis1tfs}.

If there is a bundle tensor $I^A{}_B(X)$
 that  satisfies
 \be
G_{CA}I^A{}_B=G_{BA}I^A{}_C
\ee
and is covariantly  constant
\be
\nabla_i I^A{}_B \equiv \partial _i I^A{}_B +  A_i {}^A{}_CI^C{}_B -
I^A{}_C A_i {}^C{}_B {=0}
\ee
then 
the theory has a further fermionic symmetry \cite{Hull:1985jv} of the form
\be
\label{zetran}
\delta \Psi ^A_-=\zeta _- I^A{}_B \nabla_+\Psi ^B_-
\ee
where $\zeta _-$ is a spinorial parameter.
The presence of such covariantly constant tensors reduces the structure group of the bundle.
The non-trivial  irreducible cases are the case in which the structure group is reduced to 
$U(m/2)\subset O(m)$ ($m$ even) with one matrix $I$ which is a complex structure on the fibres,  and  {the} case in which 
  the structure group is reduced to 
$Sp(m/4)\subset O(m)$ ($m/4$ integral) with three 
complex structures on the fibres satisfying the quaternion algebra, giving three extra symmetries.
However, 
the fermionic superfields vary into field equations under symmetries of the form \re{zetran}, so that they are on-shell trivial symmetries that have no Noether charge or dynamical consequences, and commute with the usual supersymmetries on-shell.
Nonetheless, such on-shell trivial transformations can be useful, as combining them with the supersymmetry transformations \re{psitrans} can under certain conditions give a supersymmetry algebra that closes off-shell
\cite{Howe:1988cj}. This will be useful here for constructing  models with $(4,0)$ supersymmetry off-shell.
For each  off-shell supersymmetery, there is a complex structure 
$J^i{}_j$ on the base and a complex structure $I^A{}_B$ on each fibre that combine to form a complex structure on the total space.

We will discuss a special class of $(4,0)$ models in  $(2,0)$ superspace and in $(4,0)$ projective superspace.  $(4,0)$ models have been discussed in $(1,0)$ superspace, in harmonic superspace and in components in, e.g.,  \cite{Howe:1988cj}, \cite{Lhallabi:1988mv} and  \cite{Gates:1994bu},\cite{Hull:1986hn}.
In the quantum theory, anomaly cancellation requires modifying the geometry, with $H$ modified by Chern-Simons terms so that $dH\propto tr(F^2-R^2)$ \cite{Hull:1985jv}.
The finiteness of (4,0) sigma models has been discussed in 
\cite{Howe:1992tg},
\cite{Sokatchev:1986an}.

\section{$(4,0)$ Off-Shell Supermultiplets}
\label{Off0}

In this section we formulate off-shell $(4,0)$ supermultiplets that generalise the  $(4,1)$ supermultiplet introduced in \cite{Hull:2016khc} and the $(4,4)$ supermultiplets of \cite{Gates:1984nk}.
We use a
 $(4,0)$ superspace
 % obtained from the $(4,1)$ superspace by omitting the minus sector. It thus has 
 with coordinates $x^\+,x^=, \theta ^+_a, \bar \theta ^{+a}$  with $\theta ^+_a$ being complex. The index $a=1,2$ is an $SU(2)$ index, 
and the two right-handed complex spinorial covariant derivatives $\bbD{+}^a$ satisfy.
\ber\nn\label{talg1}
&&\{\bbD{+a},\bbDB{+} ^b\}=~2i\delta^b_a\pa_\+~, ~~~a,b,=1,2.
\eer
 There is then a $(4,0)$ multiplet,  obtained  by truncating the  $(4,1)$ multiplet  \re {constr2},
  consisting of a pair of scalar $(4,0)$ superfields $\phi, \chi$ satisfying the constraints 
\ber\nn\label{constr20}
&&\bbDB{+}^1\phi = 0=\bbD{+2}\phi~,~~~\bbDB{+}^1 \chi =0=\bbD{+2}\chi~,\\[1mm]
&&\bbDB{+}^2\chi=-i\bbDB{+}^1\bar \phi~,~~~\bbDB{+}^2\phi=i\bbDB{+}^1\bar\chi~.
\eer
In addition we introduce a fermi multiplet consisting of a pair of spinor superfields
$\psi_-, \lambda_-$ satisfying the following constraints
\ber\nn\label{constr20f}
&&\bbDB{+}^1\psi_-= 0=\bbD{+2}\psi_- ~,~~~\bbDB{+}^1 \lambda_- =0=\bbD{+2}\lambda_-~,\\[1mm]
&&\bbDB{+}^2\lambda_-=-i\bbDB{+}^1\bar \psi_-~,~~~\bbDB{+}^2\psi_-=i\bbDB{+}^1\bar\lambda_-~.
\eer

The $(4,1)$ supermultiplet of \cite{Hull:2016khc} is very similar: there, the $(4,1)$ superfields $\hat \phi, \hat \chi$ satisfied the constraint (\ref{talg1}).
Expanding the $(4,1)$ superfields
in the extra fermionic coordinate  $\theta^-$ gives
the $(4,0)$ superfields given above.
We obtain bosonic  $(4,0)$ superfields $\phi =\hat \phi \vert _{\theta ^-=0}$ and  $\chi =\hat \chi \vert _{\theta ^-=0}$
satisfying (\ref{talg1}) together with fermionic
$(4,0)$ superfields $\psi _- =D_- \hat \phi \vert _{\theta ^-=0}$ and  $\lambda_- = D_- \hat \chi \vert _{\theta ^-=0}$
satisfying (\ref{constr20f}).
The supersymmetry transformations  that follow from the constraints above can be rewritten in $(1,0)$  superspace; see the Appendix for details.

 $(4,1)$ sigma models constructed with   $(4,1)$ superfields $\hat \phi ^i , \hat \chi ^i$  were formulated in \cite{Hull:2016khc}, giving a $4d$-dimensional geometry with coordinates
 $\hat \phi ^i \vert _{\theta =0}$ and $\hat \chi ^i \vert _{\theta =0}$, where $i=1,\dots ,d$.
 Then expanding into $(4,0)$ superspace gives coordinate superfields $\phi^i, \chi^i$ and fermionic superfields $\psi^i _-, \lambda ^i_-$ which correspond to sections of the tangent bundle of the target space (tensored with the negative chirality spinor bundle of the worldsheet).
 We will obtain more general models by allowing $\psi_-, \lambda_-$ to be sections of an arbitrary vector bundle over the target space, instead of the tangent bundle.

%\rd{ ==========================================}

%%%

\section{$(2,0)$  Superspace Formulation}

The general  $(2,0)$ sigma model action   can be written in $(2,0)$  superspace as  \cite{Dine:1986by}
\ber\label{2,0}
S= 
\int d^2x d^2 \theta
\left(k_ \alpha \partial_=\varphi^\alpha
+
\bar k_{\bar \alpha}  \partial_= \bar \varphi^{\bar \alpha} +e_{\mu \nu}\Lambda_-^{ \mu }\Lambda_-^{ \nu} + {G_{\mu \bar\nu}}\Lambda_-^{ \mu }\bar\Lambda_-^{\bar \nu}
+e_{\bar\mu \bar\nu}\bar\Lambda_-^{ \bar \mu }\bar\Lambda_-^{\bar \nu}
\right)~,
\eer
where {$G_{\mu\bar\mu}$ is} the fibre metric, and $e_{\bar\mu\bar\nu}=\overline{e_{\mu\nu}}$. 
Expanding in components or (1,0) superfields gives a Hermitian target space metric
\ber\label{GB}
g_{\alpha \bar \beta}=i(\pa_{ \alpha} \bar k_{\bar  \beta}-\pa_{\bar \beta} k_{\alpha} ) 
\eer
and
a $B$-field which, in a gauge in which $B=B^{(2,0)}+B^{(0,2)}$, has 
\ber
\label{GB2}
B^{(2,0)}_{\alpha\beta}=  i(  \pa_{\alpha} k_{\beta}     -\pa_{  \beta} k_{\alpha}   )~,
\eer  
and $B^{(0,2)}$ is the complex conjugate of this.
The fields  $\varphi^\alpha$ are $(2,0)$ chiral scalar superfields
\ber
\bbDB{+}\varphi^\alpha=0~,
\eer
and $\bar \varphi^{\bar \alpha} $ are their complex conjugates $\bar \varphi^{\bar \alpha} =
(\varphi^\alpha)^*$. 
The fields  $\Lambda^{ \mu }_-$ are $(2,0)$ fermionic chiral spinor superfields
\ber
\bbDB{+}\Lambda_-^{ \mu }=0\, ,
\eer
and $ \bar\Lambda_-^{ \bar \mu }$ are their complex conjugates.

Expanding the $(4,0)$ supermultiplets of the last section into $(2,0)$ superspace, using a similar procedure to that in \cite{Hull:2016khc}, gives $(2,0)$ superfields.
First, it gives
chiral
$(2,0)$ scalar superfields $\phi, \chi$ that transform under the extra nonmanifest supersymmetries as
\ber\label{spec0}
\bar Q_+  \phi= i\bbDB{+} \bar  {  \chi }, \qquad \bar Q_+   \chi
= -i\bbDB{+} \bar  {  \phi }
\eer
In addition it gives chiral  fermionic $(2,0)$ spinor superfields $\psi_-, \lambda_-$
 transforming under the extra nonmanifest supersymmetries as
\ber\label{spec1}
\bar Q_+  \psi_-= i\bbDB{+} \bar    \lambda_- , \qquad \bar Q_+   \lambda_-
= -i\bbDB{+} \bar    \psi_- 
\eer

The action for $d$ $(4,0)$  multiplets must take the form  \re{2,0} when written in $(2,0)$  superspace, with
$(2,0)$  chiral superfields  $\varphi^\alpha = (  \phi ^i,   \chi ^i)$ with $i=1,\dots , d$ and $\alpha =1,\dots, 2d$
and fermionic chiral superfields $\Lambda ^\mu _-= (\psi_-^a, \lambda_-^a)$ with $a=1,\dots , n$ and $\mu =1,\dots, 2n$ for some $n$.
For the complex conjugate superfields, we use the notation
$\bar \varphi^{\bar \alpha} = ( \bar \phi ^i,   \bar \chi ^i)$ and
$\bar \Lambda ^{\bar \mu} _-= (\bar\psi_-^a, \bar\lambda_-^a)$.
The transformations can then be {written as} 
\ber
\label{nonmansusy}
\bar\delta\varphi=-\bar\epsilon^+\hat \sigma_2\bbDB{+}\bar\varphi~,~~~\bar\delta\Lambda_-=-\bar\epsilon^+\hat \sigma_2\bbDB{+}\bar\Lambda_-~,
\eer
where $\hat \sigma_2$
is $\hat \sigma_2= \sigma_2 \otimes \mathbb{1}_{d\times d }$ when acting on the bosonic superfields and  
is $\hat \sigma_2= \sigma_2 \otimes \mathbb{1}_{n \times n}$
when acting on the fermionic superfields, with  $\sigma_2$ the usual Pauli matrix.

Expanding into (1,0) superspace as outlined in the Appendix gives three complex structures for the 
 target space ${\cal M}$ and three complex structures for the fibres, and both sets take the 
 form
\ber\label{comstr1}
\mathbb{I}^{(1)}=\left(\begin{array}{cc}i   &0\\
0&-i    \end{array}\right) \otimes \mathbb{1} ~,~~~
\mathbb{I}^{(2)}=\left(\begin{array}{cc}0& i \hat \sigma_2\\ i \hat \sigma_2&0\end{array}\right)~,~~~
\mathbb{I}^{(3)}=\left(\begin{array}{cc}0&- \hat \sigma_2\\
\hat \sigma_2&0\end{array}\right)~.
\eer
where 
$\mathbb{1}=
\mathbb{1}_{d\times d }$ when acting on the bosonic superfields and $\mathbb{1}=
\mathbb{1}_{n\times n }$ when acting on the fermionic superfields.
The complex structure $\mathbb{I}^{(3)}$ is the one corresponding to the supersymmetry transformations \re{nonmansusy}.
The complex structures acting on  the bosons and  the complex structures acting on  the fermions   each satisfy the quaternion algebra
\ber
\mathbb{I}^{(I)}\mathbb{I}^{(J)}=-\delta^{IJ}+\epsilon^{IJK}\mathbb{I}^{(K)}~.
\eer
For each $I$, the complex structure on ${\cal M}$ and the complex structure on the fibres together represent a complex structure on the total space.
Together the complete set defines a quaternionic structure on the total space.

The vector potential in \re{2,0} has components
$k_\alpha = (k_{\phi^i},k_{\chi^i})$ 
and the terms involving $k$ will be invariant if $k$  {satisfies}
%\footnote{ For convenience we  sometimes omit the bar on the index of the complexconjugated field.}
\ber\label{herm1}\nn
&&g_{\phi^{i}\bar\phi^{j}}-g_{\chi^{j}\bar\chi^{i}}=\bar k_{\bar\phi^i,\phi^j}-k_{\phi^j,\bar\phi^i}-\bar k_{\bar\chi^j,\chi^i}+k_{\chi^i,\bar\chi^j}=0\\[1mm]
%&&\bar k_{\bar\chi^{i},\phi^{j}}+\bar k_{\bar\phi^{i},\chi^{j}}-k_{\phi^{j},\bar\chi^{i}}-k_{\chi^{j},\bar\phi^{i}}=0~,
&&g_{\phi^{(i}\bar\chi^{j)}}=\bar k_{\bar\chi^{(i},\phi^{j)}}-k_{\phi^{(i},\bar\chi^{j)}}=0~,
\eer
which is the condition that the metric is hermitian with respect to $\mathbb{I}^{(3)}$,
as well as
\ber\nn\label{herm2}
&&\half\left(k_{\phi^{[j} ,\bar\chi^{k]}}-\bar k _{\bar\chi^{[j},\phi^{k]}}\right)_{, \bar\beta}-\bar k_{\bar\beta, \phi^{[j} \bar\chi^{k]}}=0\\[1mm]\nn
&&\half\left(\bar k_{\bar\phi^k,\phi^j}+ k_{\chi^k, \bar\chi^j}+\bar k_{\bar\chi^j,\chi^k}+ k_{\phi^j, \bar\phi^k}\right)_{, \bar\beta}
-\bar k_{\bar\beta, \phi^j\bar  \phi^k}-\bar k_{\bar\beta, \chi^k \bar\chi^j}=0\\[1mm]
&&\half\left(k_{\chi^{[j}, \bar\phi^{k]}}-\bar k _{\bar\phi^{[j}, \chi^{k]}}\right)_{,
\bar\beta}-\bar k_{\bar\beta, \chi^{[j}
\bar\phi^{k]}}=0~,
\eer
which represents the covariant constancy of the complex structures. These conditions are the same as those derived for the $(4,1)$ multiplet in \cite{Hull:2016khc}.

Turning to the fermionic part of  \re{2,0}, the variation of the terms involving $\Lambda$ gives
\ber\label{met}
G_{\psi^{(a}\bar\lambda^{b)}}=0~,~~~G_{\psi^{a}\bar\psi^{b}}-G_{\lambda^{b}\bar\lambda^{a}}=0
\eer
expressing the hermiticity of the fibre metric, and
\ber\nn\label{net}
&&G_{\psi^a\bar\nu,\bar\phi^i}-2e_{\bar\nu\bar\lambda^a,\chi^i}=0~,~~~G_{\psi^a\bar\nu,\bar\chi^i}+2e_{\bar\nu\bar\lambda^a,\phi^i}=0\\[1mm]
&&e_{\mu\nu,\chi^{[j}\bar\phi^{i]}}=0~,~~~e_{\mu\nu,\phi^{j}\bar\phi^i}+e_{\mu\nu,\chi^{i}\bar\chi^j}=0~,~~~e_{\mu\nu,\phi^{[j}\bar\chi^{i]}}=0~,
\eer
%along with their hermitean conjugate
We use the notation that subscripts  $\mu$ or $\bar \mu$ represent subscripts $\Lambda^\mu$ or $\bar\Lambda^{\bar\mu}$.

When the vector bundle is the tangent bundle, the model has $(4,1)$ supersymmetry and the complex 
 structure $\mathbb{I}^{(3)}$ on the fibres is that resulting from the complex structure on the base.
Then $G$ and $e$ are given in terms of the potential $k$ 
by $e_{\alpha \beta}=2k_{[\alpha, \beta]}$, 
$G_{\alpha\bar \beta}= k_{\alpha , \bar \beta}+\bar k_{ \bar \beta, \alpha }$
and as a result
 \re{met} and \re{net} can be shown to agree with \re{herm1} and \re{herm2}.
%We comment on this below.

\section{Projective superspace}
Projective superspace was introduced to construct actions with manifest extended supersymmetry \cite{Karlhede:1984vr} and has been studied in various dimensions; see, e.g., \cite{Lindstrom:1987ks}.
A similar but distinct approach involves harmonic superspace; see  \cite{Galperin:1984av} and references therein.

In  \cite{Hull:2016khc}, projective superspace was used to formulate models with $(4,1)$ off-shell supersymmetry.
We now adapt this construction for the $(4,0)$ models discussed in the previous sections.
$(4,0)$ projective superspace has the $(4,0)$   superspace coordinates
$x^\mu, \theta ^a $ together with the complex projective coordinate $\zeta$ on $\mathbb{CP}^1$.
We consider $(4,0)$ projective superfields
$\eta^i (x,\theta, \zeta) , \rho^a  (x,\theta, \zeta) $
 satisfying 
\ber
\na_+\eta^i=0~,~~~\breve{\na}_+\eta^i=0~,~~~\na_+\rho^a_-=0~,~~~\breve{\na}_+\rho^a_-=0~,
\eer
where 
\ber\nn\label{defN}
&&\na_+:=\bbD{+1}+\zeta\bbD{+2}~,\\[1mm]
&&\breve{\na}_+:=\bbDB{+}^1-\zeta^{-1}\bbDB{+}^2~.
\eer 
The conjugation acting on meromorphic functions of $f(\zeta)$  by
\ber
f(\zeta)\to \breve{f}(\zeta)
\eer
is
given by the composition of complex conjugation
\ber
f(\zeta)\to :f^*(\bar \zeta)\equiv ( f(\zeta))^*
\eer
and  the antipodal map
\ber
\zeta \to -\bar \zeta^{-1}
\eer
so that
\ber
\breve{f}(\zeta)
=
 f^*(-\zeta^{-1}) ~.
 \eer
 
The $(4,0)$ action is $\int[{\cal{L}}_b+{\cal{L}}_f]$, where the bosonic Lagrangian ${\cal{L}}_b$ is  formally identical to the bosonic part of the reduced $(4,1)$ action
\ber\label{4,00}
\int {\cal {L}}_b:=i\oint_C\frac{d\zeta}{2\pi i \zeta} \Delta_+\breve{\Delta}_+\left(\lambda_i(\eta,\breve{\eta})\pa_=\eta^i-\breve{\lambda}_{ i}(\eta,\breve{\eta})\pa_=\breve{\eta}^ i\right)~,
\eer
and the fermionic Lagrangian is
\ber\label{f4,00}
\int {\cal {L}}_f:=i\oint_C\frac{d\zeta}{2\pi i \zeta} \Delta_+\breve{\Delta}_+\left(
\rho^a_-h_{ab} \rho^b_-+ \rho^a_-h_{a\bar b}\breve\rho^{\bar b}_-+  \breve{\rho}^{\bar a}_-\breve{h}_{\bar a b} \rho^{b}_-+\breve{\rho}^{\bar a}_-\breve h_{\bar a\bar b}\breve{\rho}^{\bar b}_-\right)~,
\eer
where $h_{ab}$ and  $\breve{h}_{\bar a\bar b}$ are  antisymmetric and related by   conjugation
%\footnote{Note that.\rd{not sure what we want here. Cup conjugation is defined as complex conj plus..}.}. 
The metric
% formed from 
$H_{a\bar b}:= h_{a\bar b}-\breve h_{\bar b a}$ is hermitian. The contour is taken to encircle the origin and the measure is formed from two operators orthogonal to the ones in \re{defN}, and may be replaced by the $(2,0)$ measure when acting on functions of $\eta$ and $\bar\eta$:
\ber
 \Delta_+\breve{\Delta}_+\to \bbD{+}\bbDB{+} ~.
\eer

For \re{4,00}, \re{f4,00} to give the   $(2,0)$ action \re{2,0} with extended supersymmetry \re{spec0} and \re{spec1}, we choose
\ber
\label{projch}
\nn
&&\eta^i=\bar\phi^{i}+\zeta\chi^i~,~~~\bar\eta^{i}:=\breve\eta^{i}=\phi^i-\zeta^{-1}\bar\chi^{i}\\[1mm]
&&\rho^a_-=\bar\psi^{a}_-+\zeta\lambda^a_-~,~~~\bar\rho^{a}_-:=\breve\rho^{a}_-=\psi_-^a-\zeta^{-1}\bar\lambda^{a}
\eer
In the bosonic sector the metric and $B$-field is expressible in terms of the vector potentials $k_\alpha = (k_{\phi^i},k_{\chi^i})$  as in \re{GB},\re{GB2}.
These are given by
\ber\nn\label{ks}
&&k_{\phi^i}=-\oint_C\frac{d\zeta}{2\pi i \zeta}\breve{\lambda}_{i}~,~~~\bar k_{\bar\phi^i}=\oint_C\frac{d\zeta}{2\pi i \zeta}{\lambda}_i\\[1mm]
&&k_{\chi^i}=\oint_C\frac{d\zeta}{2\pi i \zeta}\zeta{\lambda}_i~,~~~~\bar k_{\bar\chi^i}=\oint_C\frac{d\zeta}{2\pi i \zeta}\zeta^{-1}\breve{\lambda}_{i}~.
\eer
The general properties
of a function $f(\eta,\breve\eta)$ ,
\ber\nn\label{fprop}
&&f_{,\phi^i }=f_{,\bar\eta^i}~,~~~~f_{,\chi^i}=\zeta f_{,\eta^i}\\[1mm]
&&f_{,\bar\phi^{i}}= f_{,\eta^{i}}~,~~~f_{, \bar \chi^{i}}=-\zeta^{-1}f_{,\bar\eta^{i}}~,
\eer
implies that the potentials satisfy
\ber\nn\label{vpots}
&&k_{\phi^i,\bar \phi^{ j}}+\bar k_{\bar\chi^i, \chi^{ j}}=0~,\\[1mm]
&&k_{\phi^i,\bar \chi^ j}-\bar k_{\bar\chi^i, \phi^j}=0~,
\eer
which is sufficient for, but does not imply, the relations \re{herm1} and \re{herm2}.  Geometrially, these conditions, together with the hermiticity \re{herm1}, imply that  for each $I$ the complex structure $\mathbb{I}^{(I)}$ is compatible with  the $B$-field
\ber\label{bcond}
(\mathbb{I}^{(I)} ) ^tB^{(1,1)}\mathbb{I}^{(I)}=B^{(1,1)}~,
\eer
where $B^{(1,1)}$
\ber
B^{(1,1)}_{\alpha\bar\alpha}=i\left(\bar k_{\bar\alpha,\alpha}+k_{\alpha,\bar\alpha}\right)~.
\eer
and
 is related to $B^{(2,0)}+B^{(0,2)}$ by a gauge transformation.
It follows that 
\ber
(\mathbb{I}^{(I)} ) ^tE\mathbb{I}^{(I)}=E~,
\eer
where
\ber
E:=g+B^{(1,1)}~.
\eer

Turning now to the fermionic sector, we find that
the fibre metric and $e$-field are given by
\ber\nn\label{gses}
&&G_{\psi^a\bar\psi^{\bar b}}=-\oint_C\frac{d\zeta}{2\pi i \zeta}H_{\bar a b}~,~~~G_{\psi^a\bar\lambda^{\bar b}}=-2\oint_C\frac{d\zeta}{2\pi i \zeta}\zeta^{-1}h_{\bar a\bar b}~,\\[1mm]\nn
&&G_{\lambda^a\bar\psi^{\bar b}}=2\oint_C\frac{d\zeta}{2\pi i \zeta}\zeta h_{ ab}~,~~~~G_{\lambda^a\bar\lambda^{\bar b}}=-\oint_C\frac{d\zeta}{2\pi i \zeta}H_{a\bar b}~,\\[1mm]\nn
&&e_{\psi^a\psi^{ b}}=\oint_C\frac{d\zeta}{2\pi i \zeta}h_{\bar a\bar b}~,~~~~~~~~~~~~e_{\lambda^a\psi^b}=\half \oint_C\frac{d\zeta}{2\pi i \zeta}\zeta H_{a\bar b}~,~~~\\[1mm]\nn
&&e_{\lambda^a\lambda^b}=\oint_C\frac{d\zeta}{2\pi i \zeta}\zeta^2 h_{ab}
~,~~~~~~~~~~e_{\bar\psi^{\bar a}\bar\psi^{\bar b}}=\oint_C\frac{d\zeta}{2\pi i \zeta}h_{ab}\\[1mm]
&&e_{\bar\psi^{\bar a}\bar\lambda^{\bar b}}=-\half   \oint_C\frac{d\zeta}{2\pi i \zeta}
\zeta^{-1}H_{a\bar b}~,~~~e_{\bar\lambda^{\bar a}\bar\lambda^{\bar b}}=\oint_C\frac{d\zeta}{2\pi i \zeta}\zeta^{-2} h_{\bar a\bar b}~.
\eer
The expressions in \re{gses} satisfy all the relations in \re{met} and \re{net}.  Proving this requires use of derivatives of the relations for a function $f(\eta,\breve\eta)$ in \re{fprop}:
\ber\nn\label{ohno}
&&f_{,\phi^i \bar \phi^{\bar j}}=f_{,\bar\eta^{\bar i}\eta^j}~,~~~f_{,\chi^i \bar \chi^{\bar j}}=-f_{,\eta^i\bar\eta^{\bar j}}\\[1mm]
&&f_{,\chi^i \bar \phi^{\bar j}}=\zeta f_{,\eta^i\eta^{j}}~,~~~f_{,\phi^i \bar \chi^{\bar j}}=-\zeta^{-1}f_{,\bar\eta^{\bar i}\bar\eta^{\bar j}}~.
\eer
This relation also means that all the geometric fields $U=(G_{\mu\bar\nu}, e_{\mu \nu}
,\bar e_{\bar\mu \bar\nu} )
$ in  \re{gses} will obey 
\ber\label{yeah}
U_{,\phi^i\bar\phi^j}+U_{,\chi^j\bar\chi^i}=0~,
\eer
in addition to \re{net}. 
It serves as a check that, in the special case  when the 
vector bundle is the tangent bundle and 
model has $(4,1)$ supersymmetry, the equation \re{yeah} follows from  \re{vpots}.

We stress that the conditions \re{yeah},  although reminicent of  the conditions \re{vpots} in the bosonic sector, are not all needed for the invariance of the fermionic part of the action, \re{met} and \re{net}:
only  \re{yeah} for $U=e_{\mu\nu}$ is required for invariance.
In section 3, we constructed the general sigma model for our off-shell $(4,0)$ multiplets.
The projective superspace action given here only gives a special sublass of these models.
Finally, we note that we made a particular choice of proejctive superfield with the ansatz \re{projch}, and 
other choices with other $\zeta $ dependence
 give a wider class of models, typically involving left- or right-moving multiplets, and/or auxiliary fields. 

\vspace{1cm}

\noindent{\bf Acknowledgements:}  This work was supported  by the EPSRC programme grant ``New Geometric
Structures from String Theory" EP/K034456/1
and STFC grant ST/L00044X/1.
UL gratefully acknowledges the hospitality of the theory group at Imperial College, London, as
well as partial financial support by the Swedish Research Council through VR grant 621-2013-4245.

\appendix
\section{$(1,0)$ superspace form of $(4,0)$ transformations}
\label{app1}

The $(4,0)$ multiplet  (\ref{talg1})
 consists of a pair of $(4,0)$ superfields $\phi, \chi$ satisfying the constraints 
 %\rd{XX}
\ber\nn\label{constr2}
&&\bbDB{+}^1\phi = 0=\bbD{+2}\phi~,~~~\bbDB{+}^1 \chi =0=\bbD{+2}\chi~,\\[1mm]
&&\bbDB{+}^2\chi=-i\bbDB{+}^1\bar \phi~,~~~\bbDB{+}^2\phi=i\bbDB{+}^1\bar\chi~.
\eer
The supersymmetry transformations can be put into the form \re{nis1tfs} by expanding in  $(1,0)$  superspace.
The $(4,0)$ multiplet in \re{constr2} can be formulated in $(1,0)$   superspace
by defining
\ber\label{1comp}
\phi \big\vert _{\theta _2^+=0,\theta _1^+=\bar \theta _1^+} = \tilde \phi , \qquad \chi \big \vert _{\theta _2^+=0,\theta _1^+=\bar \theta _1^+} = \tilde \chi
\eer
The constraints \re{constr2} then determine the terms in $\phi, \chi $ of higher order in $\theta_2,\theta _1^+-\bar \theta _1^+$
in terms of $\tilde \phi, \tilde \chi $
and give the supersymmetry transformations under the non-manifest supersymmetries.
We define four real $(4,0)$ superspace spinor derivatives $D_+$ and $\check{D}_+^{(A)}~,~A=1,2,3$ by
\ber\nn\label{8}
&&\bbD{+1}=: D_+ - i\check{D}_+^{(1)}\\[1mm]
&&\bbD{+2}=: \check{D}_+^{(2)} - \check{D}_+^{(3)}~,
\eer
Then  $D_+$ is 
the $(1,0)$ superspace spinor derivative  and the three differential operators $\check{D}_+^{(A)}~,A=1,2,3$ determine the generators of nonmanifest supersymmetries $Q_+^{(A)}$ via the constraint \re{constr2} 
\ber
\check{D}^{(A)}_+\phi\Big\vert _{\theta _1^+ = \bar \theta _1^+,\theta _2^+=0}& =&Q^{(A)}_+ \tilde \phi~,
\\
\check{D}^{(A)}_+\chi\Big\vert _{\theta _1^+ = \bar \theta _1^+,\theta _2^+=0}& =&Q^{(A)}_+ \tilde \chi~,
\eer
resulting in the following relation for the extended supersymmetries
\ber
Q_+^{(A)}\left(\begin{array}{c}\tilde\phi\\
\tilde\chi\\
\bar{\tilde\phi}\\
\bar{\tilde\chi}\end{array}\right)=:\mathbb{J}^{(A)}D_+\left(\begin{array}{c}\tilde\phi\\
\tilde\chi\\
\bar{\tilde\phi}\\
\bar{\tilde\chi}\end{array}\right)~.\eer
where the complex structures
\ber\label{comstr}
\mathbb{J}^{(A)}= \mathbb{I}^{(A)}\otimes \mathbb{1}_{d\times d}
\eer
with
\ber\label{comstr1}
\mathbb{I}^{(1)}=\left(\begin{array}{cc}i\mathbb{1}&0\\
0&-i\mathbb{1}\end{array}\right)~,~~~
\mathbb{I}^{(2)}=\left(\begin{array}{cc}0& i\sigma_2\\ i\sigma_2&0\end{array}\right)
~,~~~
\mathbb{I}^{(3)}=\left(\begin{array}{cc}0&-\sigma_2\\
\sigma_2&0\end{array}\right)
\eer
 are constant in this coordinate system and satisfy the quaternion algebra
\ber
\mathbb{J}^{(A)}\mathbb{J}^{(B)}=-\delta^{AB}+\epsilon^{ABC}\mathbb{J}^{(C)}~.
\eer
Then this gives transformations for $\tilde \phi, \tilde \chi $ of the form \re{nis1tfs}.
The (1,0) superspace formulation of the fermionic superfields can be found similarly.

%%%%%
%%%%%%%%

   \end{document}